\documentclass[review,3p]{elsarticle}

\usepackage{lineno,hyperref}
\usepackage{amssymb}
\usepackage{amsmath}
\usepackage{color}
\usepackage{soul}
\usepackage{enumerate}

\modulolinenumbers[5]

\journal{Signal Processing}

\begin{document}
	
	\begin{frontmatter}
		
		\title{A Computational Efficient Maximum Likelihood Direct Position Determination Approach for Multiple Emitters Using Angle and Doppler Measurements}

\author[mymainaddress]{Ziqiang Wang}
\ead{ziqiangwang518@std.uestc.edu.cn}
\author[sunaddress]{Yimao~Sun}
\ead{yimaosun@scu.edu.cn}		
\author[mymainaddress]{Qun~Wan*}
\ead{wanqun@uestc.edu.cn}
\cortext[mycorrespondingauthor]{Corresponding author at: School of Information and Communication Engineering, University of Electronic Science and Technology of China, Chengdu 611731, China.}
\address[mymainaddress]{School of Information and Communication Engineering, University of Electronic Science and Technology of China, Chengdu 611731, China.}
\address[sunaddress]{College of Computer Science, Sichuan University, Chengdu 610065, China.}
\author[HKUST]{Lei Xie}
\ead{eelxie@ust.hk}	
\author[NIEE]{Ning Liu}
\ead{lning1965@126.com}			
\address[HKUST]{{Department of Electronic and Computer Engineering, Hong Kong University of Science and Technology, Hong Kong.}}	
\address[NIEE]{Northern Institute of Electronic Equipment ,Beijing,100191, China.}	
		
	
		
		\begin{abstract}
			Emitter localization is widely applied in the military and civilian fields. In this paper, we tackle the problem of position estimation for multiple stationary emitters using Doppler frequency shifts and angles by moving receivers. The computational load for the exhaustive maximum likelihood (ML) direct position determination (DPD) search is insufferable.
		Based on the Pincus’ theorem and importance sampling (IS) concept, we propose a novel non-iterative ML DPD method. The proposed method transforms the original multi-dimensional grid search into random variables generation with multiple low-dimensional pseudo-probability density functions (PDF), and the circular mean is used for superior position estimation performance. The computational complexity of the proposed method is modest, and the off-grid problem that most existing DPD techniques face is significantly alleviated. Moreover, it can be implemented in parallel separately. 
		Simulation results {demonstrate} that the {proposed} ML DPD estimator can achieve better estimation accuracy than state-of-the-art DPD techniques. With a reasonable parameter choice, the estimation performance of the proposed technique is very close to the Cram\'er-Rao lower bound (CRLB), even in the adverse conditions of low signal-to-noise ratios (SNR) levels.		
		\end{abstract}
		
		\begin{keyword}
			Direct position determination\sep Maximum likelihood\sep Importance sampling\sep Monte-Carlo methods\sep Circular mean
		\end{keyword}
		
	\end{frontmatter}
	
	
	\section{Introduction}
\label{sec:Introduction}
%
%
%
%
Localization of the narrowband emitter attracts much interest in radar, sonar, satellite positioning and navigation \cite{torrieri1984statistical} \cite{yang2009approximately}. Some researchers devoted their attention to the scenario that both emitter and receivers are stationary, the angle of arrival (AOA) of
the emitter can be measured by the antenna array in the receiver \cite{sun2020eigenspace,zhang2020calibrating}. The others turned interests to scenarios that moving receivers with stationary emitter \cite{weiss2011direct} and moving receivers with moving emitter \cite{ho2007source}. The motion induces a Doppler frequency shift that is proportional to the radial velocity relatively between receiver and emitter. This additional information is the key for estimating location and velocity of the emitter in this case.


In general, two conventional processing steps are involved in traditional localization methods for a single emitter. In the first step, the receiver independently estimates parameters of interest, including direction of arrival (DOA) \cite{lin2006fsf,wang2009novel}, received signal strength
(RSS) \cite{tomic2014rss}, time of arrival (TOA) \cite{patwari2003relative} and Doppler frequency shift \cite{tahat2016look}. 
In the second step, the location of the emitter is estimated based on the intermediate parameters from the first step. Theoretically, these methods are sub-optimal since the estimation of the parameters in the first step ignores the constraint that all measurements must correspond to the same emitter location and transmitting signal. Furthermore, in the case of multiple emitters, the problem of associating estimated parameters with the corresponding emitter arises.

Different from the conventional two-step approach, a new technology called direct position determination (DPD) estimates the emitter position from received signals straightforwardly \cite{weiss2004direct,weiss2005direct,amar2008localization,oispuu2010direct,tirer2017high,qin2018decoupled,ma2019direct,zhao2020beamspace,hao2021sparse}. The DPD technology uses the data collected by multiple receivers to formulate a cost function that depends on the location of the emitter. Therefore, it inherently overcomes the parameter-emitter association problem and outperforms two-step methods in low signal-to-noise ratio (SNR) scenarios \cite{weiss2004direct}. Instead of transmitting
 the intermediate parameters, the DPD technique requires the transmission of received signals to a processing center so that higher communication bandwidth is necessary.

The maximum
likelihood (ML) estimator is well known to be an optimal technique for solving the DPD problem. However, for the scenario with multiple stationary emitters and moving receivers, the exact implementation of the ML technique requires a multi-dimensional grid search so that the real-time processing becomes impractical, since the computation resource is usually limited for moving receivers (e.g., unmanned aerial vehicles (UAVs)). To sidestep the exhaustive grid search, many DPD methods have been developed. Assuming the number of observed snapshots are infinite, \cite{weiss2005direct} proposed an approach that approximates the ML problem into multiple low-dimensional optimization problems under the condition that transmitted waveform is
known. Yet this method is not applicable for the moving receiver scenario since it is not practical for a receiver to repeat the track and visit the same places with the same velocities several times \cite{tirer2017high}. A subspace DPD algorithm was proposed in \cite{demissie2008localization} {with the assumption that} the number of emitters are known. To avoid model order determination, the minimum variance distortionless response (MVDR) concept was considered. \cite{tirer2017high,hao2019high} proposed a beamforming DPD algorithm based on Doppler frequency shift and AOA, respectively. In practice, the subspace and beamforming approaches are attractive due to the high resolution and modest computational load. However, they are sub-optimal compared to the ML estimator, and suffer from severe performance degradation for low SNR levels, the small number of snapshots and/or closely-spaced emitters, which are common in the moving receiver scenario. On the other hand, many existing ML solutions are iterative with a lower computational cost. The iterative DPD estimator introduced in \cite{oispuu2010direct} is based on the alternating projection (AP) technique. \cite{qin2018ml} proposed a decoupled ML DPD algorithm later, which is also iterative. Nevertheless, the performance of these two ML DPD estimators is closely tied to the initialization, i.e., their ML cost function will not converge
to the global maximum if initial guesses deviate significantly from true values. Moreover, the DPD estimators mentioned above select the fixed sampling position grids to serve as a possible set of candidate estimates, based on the assumption that unknown emitter locations are certainly on grids, which leads to the inevitable off-grid problem. Dense grid sampling can alleviate this problem while the cost is an excessive increase in computational complexity. Hence, a non-iterative ML estimator with rapidity needs to be studied, which should also be independent of grid density.

A Monte Carlo technique has been previously proposed for computationally efficient ML estimation of interested parameter\cite{saha2002maximum}. It builds on the global maximization theorem of Pincus \cite{pincus1968letter} and the importance sampling (IS) concept \cite{rubinstein2016simulation}. By choosing an appropriate IS
function, the multi-dimensional ML search problem can be
factorized into several low-dimensional sub-problems. This method has been applied to the frequency estimation and DOA estimation \cite{kay2000mean,wang2008importance}, in which good performance is provided. More recently, it was leveraged in joint angle and Doppler estimation \cite{wang2010maximum}, TDOA-based source localization \cite{wang2011importance}, delay estimation and joint angle and delay estimation in multi-path environments \cite{masmoudi2012maximum,IS_JADE}.

As mentioned before, it is difficult for state-of-the-art DPD techniques to combine computational load and robustness in moving receiver scenarios. Thus resorting to the IS concept is necessary. Some scholars have applied it to the DPD methods \cite{ma2020direct}. F. Ma \cite{ma2020distributed} exerted IS concept for the distributed DPD, but only one source is considered. Another IS-based DPD approach was proposed in \cite{hao2021importance} for multiple stationary emitters localization by static receivers using AOA and TOA. In this paper, we apply 
the IS technique along with the ML concept to the DPD problem 
for the scenario of moving array receivers and multiple stationary emitters, where the waveform of transmitted signals is known a priori. 
	The study of this paper is not a straightforward extension to \cite{hao2021importance}. As discussed in Section \ref{sec:Emitter location estimation}, the linear mean should be replaced by circular mean to obtain better position estimation performance in the considered scenario. We reformulate the conventional ML cost function into the form of pseudo-probability density function (PDF). 
	Thus the multi-dimensional maximization is converted to a multi-dimensional integration, which can be well approximated by IS technique and results thereby in tremendous computational savings.
	Considering the two-dimensional case that emitters are confined to a plane, the multi-dimensional optimization problem can be factorized into several two-dimensional sub-problems, or over a three-dimensional space in the spatial case. Thus the ML estimation of multiple emitter locations boils down to the computation of mean estimate from a number of easily generated realizations, multi-dimensional grid
	search is avoided. 
	The proposed algorithm considerably alleviates the off-grid problems and can be efficiently executed on multi-processor platforms in an 
	parallel computing implementation.
	
	The contributions of this manuscript are summarized as follows: (1) A computationally efficient non-iterative ML DPD algorithm for multiple emitters is proposed in moving receivers scenario, which transforms the high-dimensional estimation problem into multiple low-dimensional problems. (2) The search process is converted to the generation of samples, thus the off-grid problem is significantly alleviated. (3) The circular mean is applied for the final position estimation rather than the linear mean to enhance accuracy further.  
	
This paper is organized as follows: Section      \ref{sec:Problem} gives the signal model relating to angle and Doppler frequency shift. In Section \ref{sec:ML estimation}, Pincus’ theorem is applied for the ML estimation of positions of multiple emitters. In Section \ref{sec:Position estimation}, we derive the importance function and finish position estimation by circular mean, where the main contributions of the paper are given in Sections \ref{subsec:On the use of importance sampling function} and \ref{sec:Emitter location estimation}. Simulation results are
discussed in Section \ref{sec:Simulation}. Finally, conclusions are included in Section \ref{sec:Conclusion}
	
	We define beforehand some of the common notations that will be adopted in this work. Vectors and matrices are represented
	in lower-case and upper-case bold fonts, respectively. The Euclidean norm of
	any vector is denoted as $\left\| \cdot\right\| $, $U\left[{a,b} \right]^{\widetilde Q}$ stands for the ${\widetilde Q}$-dimensional uniform distribution between $a$ and $b$. Moreover, $\left(\cdot \right)^*$, $\left(\cdot \right)^T$ and $\left(\cdot \right)^H$ denote the conjugate, transpose and Hermitian transpose, respectively. $\otimes$ stands for the kronecker product and ${\bf{I}}_{N}$ denotes the $N \times N$ identity matrix. For a vector $\boldsymbol{\nu}$, ${\rm{diag}}\left\lbrace \boldsymbol{\nu} \right\rbrace $ is a diagonal matrix with diagonal elements be $\boldsymbol{\nu}$, and for a matrix $\bf{X}$, $\left[\bf{X} \right]_{row,col} $ denotes its $\left(row,col \right) $-th element. In addition, ${\mathbb {E}}\left\lbrace \cdot \right\rbrace $ and $\angle\left\lbrace \cdot \right\rbrace $ return the statistical expectation and complex number's phase, respectively. Finally, $j$ is the pure complex number that verifies $j^2=-1$, and $\buildrel \Delta \over =$ is used for definitions.
	\section{Problem formulation}
\label{sec:Problem}
	Consider a scenario with $L$ moving receivers, each one equipped a uniform linear array (ULA) consisting of $M$ elements, the spacing between adjacent elements is $d$. Receivers are assumed to be synchronized
	in frequency and time.
	
	Assuming that there are $Q$ stationary radio emitters , $Q$ is known a priori, the $q$-th emitter location is denoted by a $D \times 1$ vector of coordinate ${{\bf{p}}_q}$ (for
	planar geometry $D=2$ and for the spatial case $D=3$). All emitters radiate narrowband signals, the waveform and the carrier frequency $f_c$ of signals are known in advance (e.g., the training or synchronization sequences), and the bandwidth $B$ of signals
	is small compared to the inverse of the propagation time
	over array aperture. Each receiver intercepts the transmitted signals at $K$ short intervals along its trajectory. Let $D \times 1$ vector ${{\bf{u}}_{l,k}}$ and ${{\bf{v}}_{l,k}}$ ($l = \left\{ {1, 2,\ldots ,L} \right\}$, $k = \left\{ {1, 2,\ldots ,K} \right\}$) denote the position vector and velocity vector of the $l$-th receiver at the $k$-th interception interval, respectively. Hence, the complex signal vector observed by the $l$-th receiver at the $k$-th interception interval at time $t$ is given by
	\begin{equation}
		\label{eqn_r}
		{{\bf{r}}_{l,k}}(t) = \sum\limits_{q = 1}^Q {{b_{q,k,l}}} {{\bf{a}}_{l,k}}\left( {{{\bf{p}}_q}} \right){s_{q,k}}(t){e^{j2\pi {f_{q,k,l}}t}} + {{\bf{w}}_{l,k}}(t),{\mkern 1mu} {\mkern 1mu} {\mkern 1mu} {\mkern 1mu} {\mkern 1mu} 0 \le t \le T,
	\end{equation}
	where $T$ is the observation time.  ${{b_{q,k,l}}}$ is an unknown complex attenuation factor that represents the path attenuation from $q$-th emitter to the $l$-th receiver at the $k$-th interception interval. ${s_{q,k}}(t)$ is the transmitting signal from the $q$-th emitter during the $k$-th interception interval,  which is a priori knowledge. The noise vector ${{\mathbf {w}}_{l,k}}(t)\in {{\mathbb {C}}^{M\times 1}}$ is independent and normally distributed with zero mean and covariance matrix $\sigma^2{{\bf{I}}_M}$, which can be denoted as ${{\mathbf {w}}_{l,k}}(t)\;\backsim \;\mathcal {N}(\mathbf {0},\sigma ^2\mathbf {I}_{M})$. Aussuming that each interception interval is short enough and mutually exclusive, the array vectors ${{\bf{a}}_{l,k}}\left( {{{\bf{p}}_q}} \right)$ and observed frequency ${f_{q,k,l}}$ are considered quasistatic in each interval. Their expressions are
	\begin{equation}
		\label{eqn_array}
		{{\bf{a}}_{l,k}}\left( {{{\bf{p}}_q}} \right) =  {\left[ {{e^{j{{\boldsymbol{\beta }}_{l,k}^T}\left( {{{\bf{p}}_q}} \right){{\bf{d}}_{l,k,1}}}}, \ldots ,{e^{j{{\boldsymbol{\beta }}_{l,k}^T}\left( {{{\bf{p}}_q}} \right){{\bf{d}}_{l,k,M}}}}} \right]^T},
	\end{equation}
	where ${{\bf{d}}_{l,k,m}}$ denotes the position vector of $m$-th array antenna in the $l$-th receiver at the $k$-th interception interval, and
	\begin{equation}
		\label{eqn_doppler}
		{f_{q,k,l}} ={f_{c}}+{f_{c}} {\mu _{l,k}}\left( {{{\bf{p}}_q}} \right),
	\end{equation}
	where
	\begin{equation}
		\label{eqn_beta}
		{{\boldsymbol{\beta }}_{l,k}}\left( {{{\bf{p}}_q}} \right) \buildrel \Delta \over = \frac{{2\pi {f_c}}}{c}\frac{{{{\bf{u}}_{l,k}} - {{\bf{p}}_q}}}{{\left\| {{{\bf{u}}_{l,k}} - {{\bf{p}}_q}} \right\|}},
	\end{equation}
	\begin{equation}
		\label{eqn_mu_k}
		{\mu _{l,k}}\left( {{{\bf{p}}_q}} \right)\buildrel \Delta \over =\frac {\mathbf {v}_{l,k}^{{T}}({{\mathbf {p}}_{q}}-{{\mathbf {u}}_{l,k}})}{c\left \|{ {{\mathbf {p}}_{q}}-{{\mathbf {u}}_{l,k}} }\right \|},
	\end{equation}
	and $c$ is the signal’s propagation speed. More details about the observed frequency are given in \cite{amar2008localization}. As ${f_{c}}$ is known to the receiver, after down conversion, the intercepted signal frequency become $\bar {f}_{q,k,l}=f_{q,k,l}-f_{c}$.
	
	The down converted signal is sampled at $t_{n}=nT_{s}$ where $n=\{0,\ldots ,N-1\}$ and $T_{s}=T/N$. For simplicity, ${\bf {r}}_{l,k} (nT_{s})$, ${s_{q,k}}(nT_{s})$ and ${\bf {w}}_{l,k}(nT_{s})$ are represented by ${\bf {r}}_{l,k} [n]$, ${s_{q,k}}[n]$ and ${\bf {w}}_{l,k}[n]$, respectively. Then the sampled signal can be rewritten in a vector form as
	\begin{equation}
		\label{eqn_r_vec}
		{\bf {\bar r}}_{l,k} =\sum\limits_{q = 1}^Q b_{q,k,l} {\bf D}_{l,k}\left( {{{\bf{p}}_q}} \right){\bf s}_{q,k} + {\bf {\bar w}}_{l,k},
	\end{equation}
	where	
\begin{equation}
	\begin{split}
		\label{biaoshi}
		{{{\bf{\bar r}}}_{l,k}} &\buildrel \Delta \over = {\left[ {{{\bf{r}}_{l,k}^T}[0],\ldots,{{\bf{r}}_{l,k}^T}[N - 1]} \right]^T},\\
		{\bf D}_{l,k}\left( {{{\bf{p}}_q}} \right) &\buildrel \Delta \over = {{\bf{a}}_{l,k}}\left( {{{\bf{p}}_q}} \right)\otimes{\bf F}_{l,k}\left( {{{\bf{p}}_q}} \right),\\
		{{{\bf{s}}}_{q,k}} &\buildrel \Delta \over = {\left[ {{s_{q,k}}[0],\ldots,{s_{q,k}}[N - 1]} \right]^T},\\
		{{{\bf{\bar w}}}_{l,k}} &\buildrel \Delta \over = {\left[ {{{\bf{w}}_{l,k}^T}[0],\ldots,{{\bf{w}}_{l,k}^T}[N - 1]} \right]^T},\\
		{\bf F}_{l,k}\left( {{{\bf{p}}_q}} \right)&\buildrel \Delta \over ={\rm{diag}}\left\{ {1, {e^{j2\pi {f_{c}}{\mu _{l,k}}\left( {{{\bf{p}}_q}} \right)T_{s}}},\ldots ,{e^{j2\pi {f_{c}}{\mu _{l,k}}\left( {{{\bf{p}}_q}} \right)\left( N-1\right)T_{s}}}} \right\}.
	\end{split}
\end{equation}
	\section{Formulation of ML estimation}
 \label{sec:ML estimation}
	\subsection{Concentrated maximum likelihood function}
	The unknown parameters of interest are given by
\begin{equation}
	\begin{split}
		{{{\bf{\bar p}}}}&\buildrel \Delta \over ={\left[{{{\bf{p}}_1^{T}},\ldots,{{\bf{p}}_Q^{T}}} \right]^T},\\
		{{{\bf{\bar b}}}}&\buildrel \Delta \over ={\left[{{\bf{b}}_{1,1}^T,\ldots,{\bf{b}}_{K,1}^T},\ldots, {\bf{b}}_{K,L}^T\right]^T},
	\end{split}
\end{equation}
	where ${\bf{b}}_{k,l}\buildrel \Delta \over ={\left[{{b_{1,k,l}},...,{b_{q,k,l}}} \right]^T}$. We focus on the ML estimator that depends on ${{{\bf{\bar p}}}}$ and ${{{\bf{\bar b}}}}$. Since ${\bf {\bar w}}_{l,k}\;\backsim \;\mathcal {N}(\mathbf {0},\sigma ^2\mathbf {I}_{NM})$, the ML estimator coincides with the Least Squares (LS) estimator. Thus the log-likelihood function (after dropping the constant terms) can be given by
	\begin{equation}
		\label{eqn_LS}
		{L_1} =  - \frac{1}{{\sigma^2}}\sum\limits_{k = 1}^K {\sum\limits_{l = 1}^L{ {{\left\| {{{{\bf{\bar r}}}_{l,k}} - {{\bf{\bar D}}_{l,k}}{{\bf{b}}_{k,l}}} \right\|}^2}}},
	\end{equation}
	where $ {{\bf{\bar D}}_{l,k}}\buildrel \Delta \over ={\left[{\bf D}_{l,k}\left( {{{\bf{p}}_1}} \right){\bf s}_{1,k},...,{\bf D}_{l,k}\left( {{{\bf{p}}_Q}} \right){\bf s}_{Q,k} \right]}$. The path attenuation vectors that maximize \eqref{eqn_LS} are given by
	\begin{equation}
		\label{eqn_{b_k_est}}
		{{{{\bf{\hat b}}}_{k,l}}} = {\left( {{\bf{\bar D}}_{l,k}^H{{{\bf{\bar D}}}_{l,k}}} \right)^{ - 1}}{\bf{\bar D}}_{l,k}^H{\bf {\bar r}}_{l,k}.
	\end{equation}
	
	Substituting \eqref{eqn_{b_k_est}} back into \eqref{eqn_LS} yields the so-called concentrated likelihood function (CLF) which depends solely on ${{{\bf{\bar p}}}}$
	\begin{equation}
		\label{eqn_{L1}}
		{L_1} =-\frac{1}{{\sigma^2}}\left[ {\sum\limits_{k = 1}^K {\sum\limits_{l = 1}^L{{{\left\| {{{{\bf{\bar r}}}_{l,k}}} \right\|}^2} - {\bf{\bar r}}_{l,k}^H{{{\bf{\bar D}}}_{l,k}}{{\left( {{\bf{\bar D}}_{l,k}^H{{{\bf{\bar D}}}_{l,k}}} \right)}^{ - 1}}{\bf{\bar D}}_{l,k}^H{{{\bf{\bar r}}}_{l,k}}}} } \right].
	\end{equation}
	
	Instead of maximizing \eqref{eqn_{L1}}, we can equivalently maximize
	\begin{equation}
		\label{eqn_{L2}}
		{L_2} = {\sum\limits_{k = 1}^K {\sum\limits_{l = 1}^L{{\bf{\bar r}}_{l,k}^H{{{\bf{\bar D}}}_{l,k}}{{\left( {{\bf{\bar D}}_{l,k}^H{{{\bf{\bar D}}}_{l,k}}} \right)}^{ - 1}}{\bf{\bar D}}_{l,k}^H{{{\bf{\bar r}}}_{l,k}}}} }.
	\end{equation}
	
	Hence, the ML estimate of ${{{\bf{\bar p}}}}$ is then obtained from the following multi-dimensional optimization
	problem
	\begin{equation}
		\label{eqn_p_est}
		{\bf{\hat{\bar p}}} = \mathop {\arg \max \;}\limits_{{\bf{\bar p}}}{L_2}.
	\end{equation}
\subsection{Global maximization of the CLF}
A direct solution of \eqref{eqn_p_est} requires $Q \times D$-dimensional search, which is extremely computational challenging. To solve this problem, we resort to the Pincus' theorem \cite{pincus1968letter}. It provides a
mean for performing the nonlinear multi-dimensional optimization and guarantees to produce the global maximum. The Pincus' theorem states that the vector $\hat {\boldsymbol{\theta }} = [{{\hat \theta }_1},{{\hat \theta }_2}, \ldots ,{{\hat \theta }_{\widetilde Q}}]$ yields the unique global maximum of a continuous ${\widetilde Q}$-dimensional function $f\left( {\boldsymbol{\theta }} \right)$, whose ${\tilde{q}}$-th entry is given by
\begin{equation}
	\label{theta_pincus}
	{{\hat \theta }_{\tilde q}} = \mathop {\lim }\limits_{\rho  \to  + \infty } \frac{{\int  \cdots  \int {{\theta _{\tilde q}}} {e^{\rho f({\boldsymbol{\theta }})}}d{\boldsymbol{\theta }}}}{{\int  \cdots  \int {{e^{\rho f({\boldsymbol{\theta }})}}} d{\boldsymbol{\theta }}}}.
\end{equation}

Using a sufficiently large ${\rho}_0$ to replace ${\rho}$, the limit in \eqref{theta_pincus} is approximated as
\begin{equation}
	\label{theta_pincus_no_limit}
	{{\hat \theta }_{\tilde q}} \approx  \frac{{\int  \cdots  \int {{\theta _{\tilde q}}} {e^{{\rho}_0 f({\boldsymbol{\theta }})}}d{\boldsymbol{\theta }}}}{{\int  \cdots  \int {{e^{{\rho}_0 f({\boldsymbol{\theta }})}}} d{\boldsymbol{\theta }}}}.
\end{equation}

Applying \eqref{theta_pincus_no_limit} to the optimization problem \eqref{eqn_p_est} with ${\boldsymbol{\theta }} = {\bf{\bar p}}$ and $f\left( {\boldsymbol{\theta }} \right)={L_2}\left( {{\bf{\bar p}}} \right)$, and letting $D=2$ for simplifying  the exhibition, the expressions for the estimation of the ${{\bf{p}}_q}$ are given by
\begin{subequations}
	\begin{align}
		\label{P_x}
		{{\hat p}_{q,x}} &\approx \int  \cdots  \int {{p_{q,x}}} \;{\bar L _2}({\bf{\bar p}})d{\bf{\bar p}},\\
		\label{P_y}
		{{\hat p}_{q,y}} &\approx\int  \cdots  \int {{p_{q,y}}} \;{\bar L _2}({\bf{\bar p}})d{\bf{\bar p}},
	\end{align}
\end{subequations}
where ${{ p}_{q,x}}$ and ${{p}_{q,y}}$ denote x-coordinate and y-coordinate of the $q$-th emitter, respectively. ${\bar L _2}({\bf{\bar p}})$ is the normalized function of ${{e^{{\rho}_0{L_2}({\bf{\bar p}})}}}$ defined as
\begin{equation}
	\label{L_2_-}
	{\bar L _2}({\bf{\bar p}}) \buildrel \Delta \over = \frac{\displaystyle {e^{\rho _0{L_2}({\bf{\bar p}})}}}{\displaystyle \int \cdots \int {e^{\rho _0{L_2}({\bf{\bar p}})}} d{\bf{\bar p}}}.
\end{equation}

However, it is difficult to directly deal with the multi-dimensional integral \eqref{P_x} and \eqref{P_y}. Closely inspecting \eqref{L_2_-}, ${\bar L _2}({\bf{\bar p}})$ satisfies ${\displaystyle \int \cdots \int {{\bar L _2}({\bf{\bar p}})} d{\bf{\bar p}}}=1$. Considering ${\bar L _2}({\bf{\bar p}})$ as a pseudo-PDF of ${\bf{\bar p}}$, thus the estimation of ${{\bf{p}}_q}$ in \eqref{P_x} and \eqref{P_y} can be alternatively regarded as statistical expectations
\begin{equation}
	\label{E_p_xy}
	{{\hat p}_{q,x}} = {{\mathbb {E}}_{{\bf{\bar p}}}}\left\{ {{p_{q,x}}} \right\}\;\;\;\;{\rm{and}}\;\;\;\;{{\hat p}_{q,y}} = {{\mathbb {E}}_{{\bf{\bar p}}}}\left\{ {{p_{q,y}}} \right\}.
\end{equation}

Intuitively, we can approximate the expectations in \eqref{E_p_xy} with the generation of $R$ realizations $\left\{ {{{{\bf{\bar p}}}^{(r)}}} \right\}_{r = 1}^R$ distributed according to ${\bar L _2}({\bf{\bar p}})$,
\begin{equation}
	\label{average_p_xy}
	{{\hat p}_{q,x}} = \frac{1}{R}\sum\limits_{r = 1}^R {p_{q,x}^{(r)}} \;\;\;\;\;{\rm{and}}\;\;\;\;{{\hat p}_{q,y}} = \frac{1}{R}\sum\limits_{r = 1}^R {p_{q,y}^{(r)}}.
\end{equation}

In this way, the complicated integrations can be approximated by
the sample mean estimates. Clearly, according to the law of
large numbers \cite{Intuitive_probability}, the sample mean estimate will converge to the corresponding expectation as $R \to \infty $. Thus the increase of $R$ results in the decrease of the variance of ${{\hat p}_{q,x}}$ and ${{\hat p}_{q,y}}$. Unfortunately, ${\bar L _2}({\bf{\bar p}})$ is still highly non-linear and cannot be practically used to generate $\left\{ {{{{\bf{\bar p}}}^{(r)}}} \right\}_{r = 1}^R$. We shall resort to the IS concept and approximately find a simple PDF for realization generation.
	\section{Position estimation using importance sampling}
		\label{sec:Position estimation}
\subsection{Utilization of the importance sampling function}
\label{subsec:On the use of importance sampling function}
Equations \eqref{P_x} and \eqref{P_y} are equivalent to the following forms respectively
\begin{subequations}
	\begin{align}
		\label{P_x_im}
		{{\hat p}_{q,x}} &= \int  \cdots  \int {{p_{q,x}}} \;\frac{{{{\bar L}_2}({\bf{\bar p}})}}{{{{\bar L}_{im}}({\bf{\bar p}})}}{{\bar L}_{im}}({\bf{\bar p}})d{\bf{\bar p}},\\
		\label{P_y_im}
		{{\hat p}_{q,y}} &= \int  \cdots  \int {{p_{q,y}}} \;\frac{{{{\bar L}_2}({\bf{\bar p}})}}{{{{\bar L}_{im}}({\bf{\bar p}})}}{{\bar L}_{im}}({\bf{\bar p}})d{\bf{\bar p}},
	\end{align}
\end{subequations}
where ${{{{\bar L}_{im}}({\bf{\bar p}})}}$ is the other pseudo-PDF called importance function \cite{kay2000mean} \cite{doucetsequential}. ${{{{\bar L}_{im}}({\bf{\bar p}})}}$ is generally designed as a simple function of ${\bf{\bar p}}$ for easily generating realizations $\left\{ {{{{\bf{\bar p}}}^{(r)}}} \right\}_{r = 1}^R$, which is also desired as similar as possible to ${{{{\bar L}_{2}}({\bf{\bar p}})}}$. Thus \eqref{P_x_im} and \eqref{P_y_im} are regarded as computing expectation of transformed random variables
\begin{equation}
	\label{E_p_xy_tr}
	{{\hat p}_{q,x}} = {{\mathbb {E}}_{{\bf{\bar p}}}}\left\{ {\eta \left( {{\bf{\bar p}}} \right){p_{q,x}}} \right\}\;\;\;{\rm{and}}\;\;\;{{\hat p}_{q,y}} = {{\mathbb {E}}_{{\bf{\bar p}}}}\left\{ {\eta \left( {{\bf{\bar p}}} \right){p_{q,y}}} \right\},
\end{equation}
where
\begin{equation}
	\label{E_eta}
	\eta \left( {{\bf{\bar p}}} \right)\; \buildrel \Delta \over = \;\frac{{{{\bar L}_2}\left( {{\bf{\bar p}}} \right)}}{{{{\bar L}_{im}}\left( {{\bf{\bar p}}} \right)}}.
\end{equation}

Now we turn into the approximate choice for $\bar{L}_{im}(\bar{\mathbf{p}})$. In
order to generate $\left\{ {{{{\bf{\bar p}}}^{(r)}}} \right\}_{r = 1}^R$ easily, intuitively ${{{{\bar L}_{im}}({\bf{\bar p}})}}$ shall be designed separable in terms of the $Q$ emitter locations
\begin{equation}
	\label{L_im_-seperate}
	{{\bar {L}}_{im}}\left( {{\bf{\bar p}}} \right) = \prod\limits_{q = 1}^Q {{{\bar {\mathcal L}}_q}\left( {{{\bf{p}}_q}} \right)},
\end{equation}
where ${{{\bar {\mathcal L}}_q}\left( {{{\bf{p}}_q}} \right)}$ is a pseudo-PDF of ${{{\bf{p}}_q}}$. ${{{{\bar L}_{im}}({\bf{\bar p}})}}$ in form \eqref{L_im_-seperate} can be regarded as a joint pseudo-PDF corresponding to multiple mutually independent random variables, which can be easily generated using $\left\{{{{\bar {\mathcal L}}_q}\left( {{{\bf{p}}_q}} \right)}\right\}_{q=1}^Q$. Hence, considering the design of ${{{{\bar L}_{im}}({\bf{\bar p}})}}$ shall be appropriate
approximation of ${{{{\bar L}_2}\left( {{\bf{\bar p}}} \right)}}$ satisfies \eqref{L_im_-seperate}.

Revisiting \eqref{eqn_{L2}}, ${L_2}\left( {\bf{\bar p}} \right)$ can be approximated as a separable function under the condition that ${{\bf{\bar D}}_{l,k}^H{{{\bf{\bar D}}}_{l,k}}}$ is replaced by an invertible diagonal matrix. It can be observed that the entry in the $i$-th row and the $g$-th column of matrix ${{\bf{\bar D}}_{l,k}^H{{{\bf{\bar D}}}_{l,k}}}$ can be expressed as
\begin{equation}
	\begin{split}
		\label{D_k^hD_k}
		\left[{{\bf{\bar D}}_{l,k}^H{{{\bf{\bar D}}}_{l,k}}}\right]_{i,g} = &\sum\limits_{n = 0}^{N - 1}{s_{i,k}^ * \left( {n{T_s}} \right){s_{g,k}}\left( {n{T_s}} \right){e^{j2\pi {f_c}\left[ {{\mu _{l,k}}\left( {{{\bf{p}}_{g}}} \right) - {\mu _{l,k}}\left( {{{\bf{p}}_i}} \right)} \right]n{T_s}}}}\\ &\times {\sum\limits_{m = 1}^M {{e^{j\left[ {{\bf{\beta }}_{l,k}^T\left( {{{\bf{p}}_{g}}} \right) - {\bf{\beta }}_{l,k}^T\left( {{{\bf{p}}_i}} \right)} \right]{{\bf{d}}_{k,m}}}}}}{\mkern 1mu} {\mkern 1mu} {\mkern 1mu} {\mkern 1mu} {\mkern 1mu}{\mkern 1mu}{\mkern 1mu}{\mkern 1mu}{\mkern 1mu}{\mkern 1mu}{\mkern 1mu}{\mkern 1mu}{\mkern 1mu}{\mkern 1mu}{\mkern 1mu}{\mkern 1mu}{\mkern 1mu}{\mkern 1mu}{\mkern 1mu}{\mkern 1mu}i,g = 1,2, \ldots ,Q.
	\end{split}
\end{equation}
The diagonal elements are
\begin{equation}
	\label{D_k^hD_k_dig}
	\left[{{\bf{\bar D}}_{l,k}^H{{{\bf{\bar D}}}_{l,k}}}\right]_{i,i} =M{\left\| {{{\bf{s}}_{i,k}}} \right\|^2}, i=1,\ldots,Q.
\end{equation}

We verify statistically that the off-diagonal entries of ${{\bf{\bar D}}_{l,k}^H{{{\bf{\bar D}}}_{l,k}}}$ are much smaller compared to $\left[{{\bf{\bar D}}_{l,k}^H{{{\bf{\bar D}}}_{l,k}}}\right]_{i,i}$ with a high probability for almost all possible emitter loactions. To this end, let us define
\begin{equation}
	\begin{split}
		\label{ratio}
		{\delta _{i,g}} \buildrel \Delta \over = &\frac{{ {\left|\sum\limits_{n = 0}^{N - 1} {s_{i,k}^*\left( {n{T_s}} \right){s_{g,k}}\left( {n{T_s}} \right){e^{j2\pi {f_c}\left[ {{\mu _k}\left( {{{\bf{p}}_{g}}} \right) - {\mu _k}\left( {{{\bf{p}}_i}} \right)} \right]n{T_s}}}}\right| } }}{{M{{\left\| {{{\bf{s}}_{i,k}}} \right\|}^2}}}\\
		&\times \left|{\sum\limits_{m = 1}^M {{e^{j\left[ {{\bf{\beta }}_k^T\left( {{{\bf{p}}_{g}}} \right) - {\bf{\beta }}_k^T\left( {{{\bf{p}}_i}} \right)} \right]{{\bf{d}}_{k,m}}}}} }\right|\qquad i\ne g,
	\end{split}
\end{equation}
as the ratio of the off-diagonal entries over diagonal entries of ${{\bf{\bar D}}_{l,k}^H{{{\bf{\bar D}}}_{l,k}}}$. Then, a large number of random variable vectors ${{{\bf{p}}_i}}$ and ${{{\bf{p}}_{g}}}$ are generated, which are independent identically
distributed (IID) in $U\left[{-100,100} \right]^2$ Km, to compute the complementary cumulative distribution function (CCDF) of ${\delta _{i,g}}$.
\begin{figure}[!t]
	\centering
	\includegraphics[width=1\linewidth]{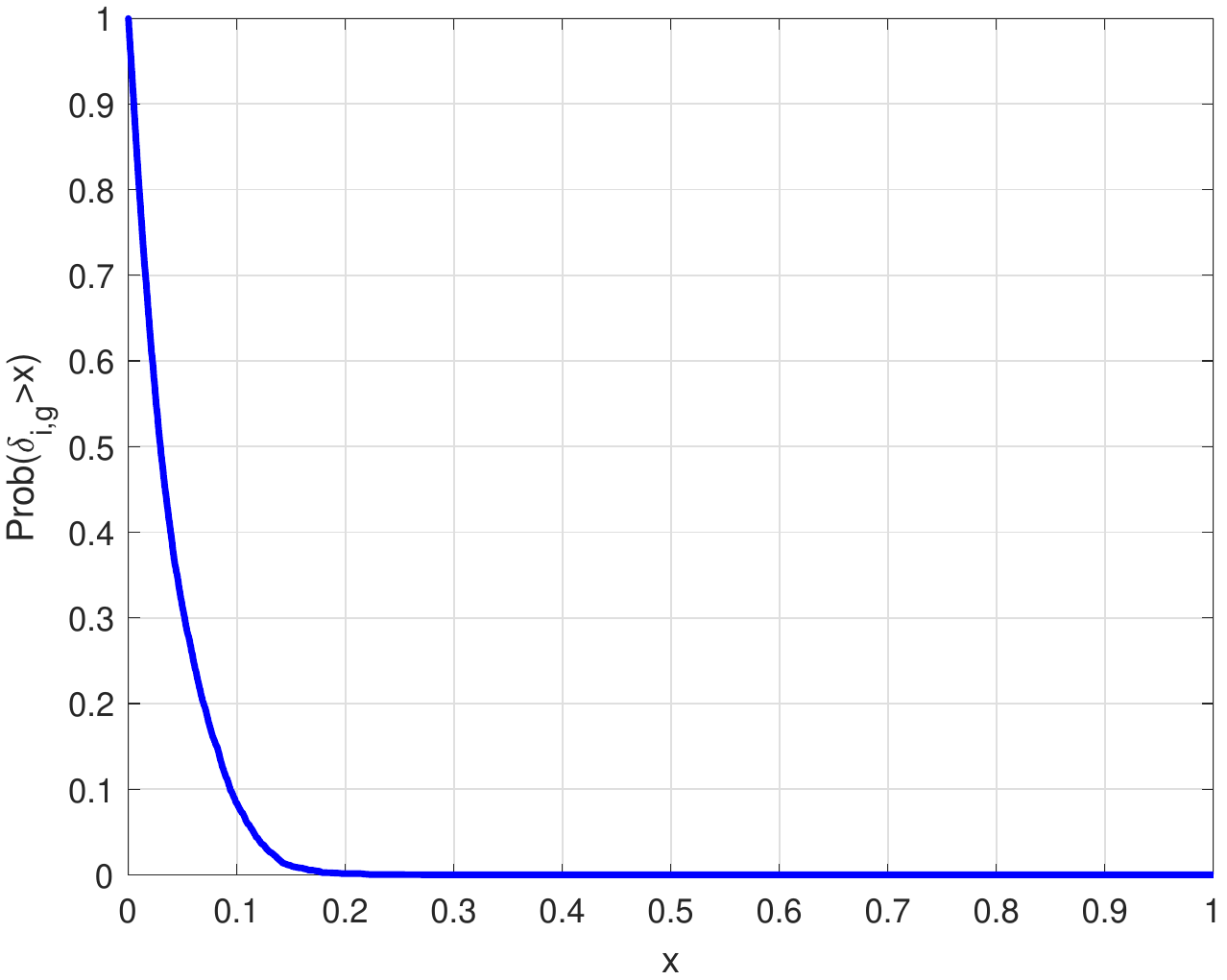}
	\caption{CCDF of the ${\delta _{i,g}}$. The length of the $k$-th interception interval is $T=12.8$ms with $N=100$ samples. The receiver is in the origin of coordinate system with 300m/s velocity, equipped with $M=3$ half wavelength spaced antennas.}
	\label{fig_ratio}
\end{figure}

As shown in Fig.\ref{fig_ratio}, $\left[{{\bf{\bar D}}_{l,k}^H{{{\bf{\bar D}}}_{l,k}}}\right]_{i,i}$ is indeed dominant compared to corresponding off-diagonal elements, since ${\delta _{i,g}}$ almost has a zero probability to exceed 0.15. Therefore, a valid approximation for \eqref{D_k^hD_k} can be obtained
\begin{equation}
	\label{D_k^hD_k_approximation}
	{{\bf{\bar D}}_{l,k}^H{{{\bf{\bar D}}}_{l,k}}}\approx M{\rm{diag}}\left\{ {{{\left\| {{{\bf{s}}_{1,k}}} \right\|}^2}, {{\left\| {{{\bf{s}}_{2,k}}} \right\|}^2},\ldots ,{{\left\| {{{\bf{s}}_{Q,k}}} \right\|}^2}} \right\},
\end{equation}
Substituting \eqref{D_k^hD_k_approximation} into \eqref{eqn_{L2}}, ${L_2}$ is approximately in proportion to the
superposition of $Q$ separated terms
\begin{equation}
	\label{eqn_L2_appro}
	{L_2}\left( {{\bf{\bar p}}} \right) \approx \frac{1}{M}\sum\limits_{q = 1}^Q {{I_q}\left( {{{\bf{p}}_q}}\right)  },
\end{equation}
where
\begin{equation}
	\label{I_Q}
	{I_q}\left( {{{\bf{p}}_q}}\right) = \sum\limits_{k = 1}^K {\sum\limits_{l = 1}^L {\frac{{{{\left|  {{\bf{\bar r}}_{l,k}^H{{\bf{D}}_{l,k}}\left( {{{\bf{p}}_q}} \right){{\bf{s}}_{q,k}}}\right| }^2}}}{{{{\left\| {{{\bf{s}}_{q,k}}} \right\|}^2}}}}}.
\end{equation}

Then, the importance function \eqref{L_im_-seperate} is transformed to
\begin{equation}
	\label{L_im_def}
	{{\bar {L}}_{im}}\left( {{\bf{\bar p}}} \right)= \frac{\displaystyle {e^{\rho _1\sum\nolimits_{q = 1}^Q  {{I_q}\left( {{{\bf{p}}_q}}\right)}}}}{\displaystyle \int \cdots \int {e^{\rho _1\sum\nolimits_{q = 1}^Q  {{I_q}\left( {{{\bf{p}}_q}}\right)}}} d{\bf{\bar p}}},
\end{equation}
where ${\rho _1}$ is the new design parameter involving factor $\frac{1}{M}$. It should be noted that the choices of ${\rho _0}$ and ${\rho _1}$ affect the performance of the proposed estimator. ${{\bar {L}}_{im}}\left( {{\bf{\bar p}}} \right)$ can indeed be factorized in terms of the $Q$ emitter locations as \eqref{L_im_-seperate}, where ${{{\bar {\mathcal L}}_q}\left( {{{\bf{p}}}} \right)}$ is expressed as
\begin{equation}
	\label{L_im_q_exp}
	{\bar {\cal L} _q}\left( {\bf{p}} \right) = \frac{\displaystyle{{e^{{\rho _1} {{I_q}\left( {\bf{p}} \right)} }}}}{{\displaystyle\int {\int {{e^{{\rho _1}{{I_q}\left( {{\bf{p}}} \right)} }}} d{\bf{p}}} }}.
\end{equation}

Thus the generation of ${{{{\bf{\bar p}}}^{(r)}}}$ can be approximated as generating $\left\{ {{{{\bf{p}}}^{(r)}_q}} \right\}_{q = 1}^Q$ with bivariate distribution \eqref{L_im_q_exp}, which can be implemented
separately and run in parallel with a much faster and less
complex execution.

One specific way for generating required vector realizations is based on the inverse probability transformation applied in \cite{Intuitive_probability}. Let ${F_{\bf{X}}}\left( {\bf{x}} \right)$ be the cumulative distribution function (CDF) of random variable vector ${\bf{X}}$ and generating $R$ IID uniform random numbers, i.e., $\left\{ {{{{u}}^{(r)}}} \right\}_{r = 1}^R\in U\left[{0,1} \right]$. Thus  the variable vector realization ${{{{{\bf{x}}}}^{(r)}}}$ can be generated as
\begin{equation}
	\label{x_r}
	{{{{{\bf{x}}}}^{(r)}}}=\mathop {\arg \min }\limits_{\bf{x}} \left| {{F_{\bf{X}}}\left( {\bf{x}} \right)-{{{{u}}^{(r)}}}} \right|	.
\end{equation}

However, this way still requires a two-dimensional search process for generating ${{{{\bf{p}}}^{(r)}_q}}$. Recalling \eqref{L_im_q_exp} that depicts a joint distribution, thus ${\bar {\cal L} _q}\left( {\bf{p}} \right)$ can be factorized as the product of a marginal PDF and a conditional PDF as follows
\begin{subequations}
	\begin{align}
		\label{mpdf-cpdf1}
		{\bar {\cal L} _q}\left( {p_x,p_y} \right)&={\bar {\cal L} _{p_{q,x}}}\left( {p_x} \right){\bar {\cal L} _{{p_{q,y}}\left| {{p_{q,x}}} \right.}}\left( {{p_{y}}\left| {{p_{x}}} \right.} \right),\\
		\label{mpdf-cpdf2}
		{\bar {\cal L} _q}\left( {p_x,p_y} \right)&={\bar {\cal L} _{p_{q,y}}}\left( {p_y} \right){\bar {\cal L} _{{p_{q,x}}\left| {{p_{q,y}}} \right.}}\left( {{p_{x}}\left| {{p_{y}}} \right.} \right),
	\end{align}
\end{subequations}
where ${\bar {\cal L} _{p_{q,x}}}\left( {p_x} \right)$ is the marginal PDF of x-coordinate $p_{q,x}$ and ${\bar {\cal L} _{{p_{q,y}}\left| {{p_{q,x}}} \right.}}\left( {{p_{y}}\left| {{p_{x}}} \right.} \right)$ is the conditional PDF of y-coordinate $p_{q,y}$ given $p_{q,x}$. The definition of ${\bar {\cal L} _{p_{q,y}}}\left( {p_y} \right)$ and ${\bar {\cal L} _{{p_{q,x}}\left| {{p_{q,y}}} \right.}}\left( {{p_{x}}\left| {{p_{y}}} \right.} \right)$ is in the same manner. ${\bar {\cal L} _{p_{q,x}}}\left( {p_x} \right)$ can be given by
\begin{equation}
	\label{mpdf-px}
	{\bar {\cal L} _{p_{q,x}}}\left( {p_x} \right)=\int{{\bar {\cal L} _q}\left( {p_x,p_y} \right)d{p_y}},
\end{equation}
which is used to generate the realization ${{{{{p}}}_{q,x}^{(r)}}}$. Then the conditional PDF of ${{p_{q,y}}}$ given ${{{{{p}}}_{q,x}^{(r)}}}$ can be expressed as following
\begin{equation}
	\label{cpdf-py|px}
	{\bar {\cal L} _{{p_{q,y}}\left| {{p_{q,x}}} \right.}}\left( {{p_{y}}\left| p_x={{{{{p}}}_{q,x}^{(r)}}}\right.} \right)=\frac{{\bar {\cal L} _q}\left( {{{{{{p}}}_{q,x}^{(r)}}},p_y} \right)}{{\bar {\cal L} _{p_{q,x}}}\left( {{{{{p}}}_{q,x}^{(r)}}} \right)}.
\end{equation}

By \eqref{cpdf-py|px} the realization ${{{{{p}}}_{q,y}^{(r)}}}$ is generated,  thus the two-dimensional search process for generating ${{{{\bf{p}}}^{(r)}_q}}=[{{{{{p}}}_{q,x}^{(r)}}}, {{{{{p}}}_{q,y}^{(r)}}}]^T $ is converted to two line search. We have just showed the process for generating ${{{{{p}}}_{q,x}^{(r)}}}$ firstly and then ${{{{{p}}}_{q,y}^{(r)}}}$. The generation of ${{{{{p}}}_{q,y}^{(r)}}}$ using ${\bar {\cal L} _{p_{q,y}}}\left( {p_y} \right)$ and then ${{{{{p}}}_{q,x}^{(r)}}}$ with ${\bar {\cal L} _{{p_{q,x}}\left| {{p_{q,y}}} \right.}}\left( {{p_{x}}\left| p_y={{{{{p}}}_{q,y}^{(r)}}}\right.} \right)$ is straightforward. 
\subsection{Emitter location estimation}
\label{sec:Emitter location estimation} 
Recalling the expression of ${{\hat p}_{q,x}}$ and ${{\hat p}_{q,y}}$ in \eqref{E_p_xy_tr} and using the IS function defined in \eqref{L_im_def}, the position estimator of the $q$-th emitter can be equivalent to estimate the linear mean \cite{hao2021importance} of generated realizations
\begin{equation}
	\label{average_p_xy_q}
	{{\hat p}_{q,x}} = \frac{1}{R}\sum\limits_{r = 1}^R {p_{q,x}^{(r)}\eta \left({{{{\bf{\bar p}}}^{(r)}}} \right)} \;\;\;\;\;\;{\rm{and}}\;\;\;\;{{\hat p}_{q,y}} = \frac{1}{R}\sum\limits_{r = 1}^R {p_{q,y}^{(r)}\eta \left({{{{\bf{\bar p}}}^{(r)}}} \right)},
\end{equation}
where 
\begin{equation}
	\label{E_eta_explict}
	\eta \left({{{{\bf{\bar p}}}^{(r)}}} \right)=\frac{\displaystyle {e^{\rho _0{L_2}\left({{{{\bf{\bar p}}}^{(r)}}}\right)}{\displaystyle \int \cdots \int {e^{\rho _1\sum\nolimits_{q = 1}^Q  {{I_q}\left( {{{\bf{p}}_q}}\right)}}} d{\bf{\bar p}}}}}{e^{\rho _1\sum\nolimits_{q = 1}^Q  {{I_q}\left(p_{q,x}^{(r)},p_{q,y}^{(r)} \right)}}\displaystyle \int \cdots \int {e^{\rho _0{L_2}({\bf{\bar p}})}} d{\bf{\bar p}}}.
\end{equation}

However, the linear mean just averages all realizations without considering the existence of outlier seeds, which may result in the inevitable estimation bias. An improvement is using the circular mean \cite{mardia1972statistics} instead. As mentioned in \cite{IS_JADE}, under the condition that ${\rho _0}$ is as high as desired, the circular mean estimation corresponds to the realization that minimizes the Euclidean distance to the true parameter. For a given random variable $Z \in \left[ { - \pi ,\pi } \right]$, let ${h}\left( {Z} \right)$ be a transformation of $Z$, then the circular mean of ${h}\left( {Z} \right)$ can be expressed as
\begin{equation}
	\label{circular mean def}
	{\hat Z_c} = \angle \;\frac{1}{R}\sum\limits_{r = 1}^R h ({Z^{(r)}}){e^{j{Z^{(r)}}}},
\end{equation}
where $\left\{ {{{Z}^{(r)}}} \right\}_{r = 1}^R$ are the realizations of random variable $Z$. 

Assuming that the x-coordinates and y-coordinates of all emitters are in the range $\left[ { p_{x,min} , p_{x,max} } \right]$ and $\left[ { p_{y,min} , p_{y,max} } \right]$, respectively. Then the alternative formulation of the estimators shown in \eqref{E_p_xy_tr} using circular mean are expressed as
\begin{subequations}
	\begin{align}
		\label{P-q_x_est_final}
		{{\hat p}_{q,x}} &= {d_{x}}\left[\frac{1}{2\pi}\angle \;\frac{1}{R}\sum\limits_{r = 1}^R {e^{j2\pi\left(-\frac{1}{2}+\frac{{p_{q,x}^{(r)}}-p_{x,min}}{{d_{x}}}\right)}}{\eta \left({{{{\bf{\bar p}}}^{(r)}}} \right)}+\frac{1}{2}\right]+p_{x,min},\\
		\label{P-q_y_est_final}
		{{\hat p}_{q,y}} &= {d_{y}}\left[\frac{1}{2\pi}\angle \;\frac{1}{R}\sum\limits_{r = 1}^R {e^{j2\pi\left(-\frac{1}{2}+\frac{{p_{q,y}^{(r)}}-p_{y,min}}{{d_{y}}}\right)}}{\eta \left({{{{\bf{\bar p}}}^{(r)}}} \right)}+\frac{1}{2}\right]+p_{y,min},
	\end{align}
\end{subequations}
where 
\begin{equation}
	{d_{x}}=p_{x,max}-p_{x,min} \;\;\;\;{\rm{and}}\;\;\;\;{d_{y}}=p_{y,max}-p_{y,min}.\nonumber
\end{equation}

Note the normalization
integral term in $\eta \left({{{{\bf{\bar p}}}^{(r)}}} \right)$ will not affect the result of $\angle \;\left\lbrace   \cdot  \right\rbrace $ in \eqref{P-q_x_est_final} and \eqref{P-q_y_est_final}, which can be omitted to reduce the computational
load. Actually, defining the weighting coefficient as
\begin{equation}
	\begin{split}
		\label{E_eta_another_explict}
		\eta '\left({{{{\bf{\bar p}}}^{(r)}}} \right)=\exp \left\{{\rho _0{L_2}\left({{{{\bf{\bar p}}}^{(r)}}}\right)
			-\rho _1\sum\nolimits_{q = 1}^Q  {{I_q}\left(p_{q,x}^{(r)},p_{q,y}^{(r)} \right)}}\right.\\
		\left. {-\mathop {\max }\limits_{1 \le r \le R} \left( {\rho _0{L_2}\left({{{{\bf{\bar p}}}^{(r)}}}\right)-\rho _1\sum\nolimits_{q = 1}^Q  {{I_q}\left(p_{q,x}^{(r)},p_{q,y}^{(r)} \right)}} \right)}\right\},
	\end{split}
\end{equation}
and replacing $\eta \left({{{{\bf{\bar p}}}^{(r)}}} \right)$ with $\eta '\left({{{{\bf{\bar p}}}^{(r)}}} \right)$ in \eqref{P-q_x_est_final} and \eqref{P-q_y_est_final} can further reduce the computational consumption and guarantee the accuracy \cite{wang2008importance}.
\subsection{Implementation details}	
\label{sec:Implementation details}
For the sake of clarity, we give all the necessary details for the location estimation of all $Q$ emitters by importance sampling as follows:	
\begin{enumerate}[STEP 1:]
	\item Derive separable term $	{I_q}\left( {{{\bf{p}}}}\right)$ by \eqref{I_Q} for every $q = 1,2, \ldots ,Q$ with the complex signal vectors $\left\{ {{{\bf{\bar r}}}_{l,k}} \right\}_{l=1,k = 1}^{L,K}$ observed from all interception intervals.
	\item \label{step_be}Choose ${N_x} \times {N_y}$ grid points in coordinate range $\left[ { p_{x,min} , p_{x,max} } \right] \times\left[ { p_{y,min} , p_{y,max} } \right]$. All grid points $\left( {{p_{{n_x}}},{p_{{n_y}}}} \right)$ have discretization steps $\Delta _{p_x}$ and $\Delta _{p_y}$ ($n_x\in \left\lbrace  {1,2, \ldots ,N_x} \right\rbrace , n_y\in \left\lbrace {1,2, \ldots ,N_y}\right\rbrace  $). Then, by approximating integrals with discrete sums, the pseudo-PDF ${\bar {\cal L} _q}\left( {\bf{p}} \right)$ for every $q$ can be approximated as
	\begin{equation}
		\label{l_q_appromax}
		{\bar {\cal L} _q}\left( {{p_{{n_x}}},{p_{{n_y}}}} \right)\approx \frac{{{e^{{\rho _1}{I_q}\left( {{p_{{n_x}}},{p_{{n_y}}}} \right)}}}}{{\sum\nolimits_{{n_x} = 1}^{{N_x}} {\sum\nolimits_{{n_y} = 1}^{{N_y}} {{e^{{\rho _1}{I_q}\left( {{p_{{n_x}}},{p_{{n_y}}}} \right)}}\Delta _{{p_x}}\Delta _{{p_y}}} } }}.
	\end{equation}
	\item\label{step_mid} Compute the marginal PDF ${\bar {\cal L} _{p_{q,x}}}\left( {p_x} \right)$ from \eqref{l_q_appromax}
	\begin{equation}
		\label{l_q_x_appromax}
		{\bar {\cal L} _{p_{q,x}}}\left( {p_{{n_x}}} \right) = {{\sum\nolimits_{{n_y} = 1}^{{N_y}} {{\bar {\cal L} _q}\left( {{p_{{n_x}}},{p_{{n_y}}}} \right)\Delta _{p_y}} }},
	\end{equation}
	then compute the corresponding CDF
	\begin{equation}
		\label{l_q_x_cdf}
		{\bar {\cal G} _{p_{q,x}}}\left( {p_{{n_x}}} \right) = {{\sum\nolimits_{{n'} = 1}^{{n_x}} {{\bar {\cal L} _{p_{q,x}}}\left( {p_{{n'}}} \right)\Delta _{p_x}} }}.
	\end{equation}
	\item Generate $R$ uniform random numbers $\left\{ {{{u}^{(r)}_{q,x}}} \right\}_{r = 1}^R\sim U\left[{0,1} \right]^1$ and perform linear interpolation to $	{\bar {\cal G} _{p_{q,x}}}\left( {p_{{n_x}}} \right) $ for the generation of $R$
	position x-coordinate realizations 
	\begin{equation}
		\label{prqx}
		{p_{q,x}^{(r)}}=\mathop {\arg \min }\limits_{ {p_{{x}}}} \left| {{\bar {\cal G} _{p_{q,x}}}\left( {p_{{x}}} \right)-{{{u}^{(r)}_{q,x}}}} \right|	.
	\end{equation}
	
	\item Compute the the conditional PDF of ${{p_{q,y}}}$ given ${{{{{p}}}_{q,x}^{(r)}}}$ with linear interpolation
	\begin{equation}
		\label{cpdf-py|px-appro}
		{\bar {\cal L} _{{p_{q,y}}\left| {{p_{q,x}}} \right.}}\left( {{p_{{n_y}}}\left| p_x={{{{{p}}}_{q,x}^{(r)}}}\right.} \right)=\frac{{\bar {\cal L} _q}\left( {{{{{{p}}}_{q,x}^{(r)}}},{p_{{n_y}}}} \right)}{	{\bar {\cal L} _{p_{q,x}}}\left( {{{{{p}}}_{q,x}^{(r)}}} \right) }.
	\end{equation}
	\item Evaluate the CDF $	{\bar {\cal G} _{{p_{q,y}}\left| {{p_{q,x}}} \right.}}\left( {{p_{{n_y}}}\left| {{{{{p}}}_{q,x}^{(r)}}}\right.} \right)$ as similarly as \eqref{l_q_x_cdf} and generate uniform random numbers $\left\{ {{{u}^{(r)}_{q,y}}} \right\}_{r = 1}^R\sim U\left[{0,1} \right]^1$.
	\item\label{step_ed}Similarly as \eqref{prqx}, obtain 
	position y-coordinate realizations $\left\lbrace {p_{q,y}^{(r)}}\right\rbrace_{r=1}^R $ with ${\bar {\cal G} _{{p_{q,y}}\left| {{p_{q,x}}} \right.}}\left( {{p_{{y}}}\left| {{{{{p}}}_{q,x}^{(r)}}}\right.} \right)$, using linear interpolation technique as well.
	\item Repeat STEP \ref{step_be}-STEP \ref{step_ed} until finishing the generation of $\left\lbrace {{{{\bf{\bar p}}}^{(r)}}}\right\rbrace _{r=1}^R$ for all $Q$ emitters.
	\item Finally estimate the location for the $q$-th emitter  \end{enumerate}
\begin{subequations}
	\begin{align}
		\label{P-q_x_est_final_fin}
		{{\hat p}_{q,x}}& = {d_{x}}\left[\frac{1}{2\pi}\angle \;\frac{1}{R}\sum\limits_{r = 1}^R {e^{j2\pi\left(-\frac{1}{2}+\frac{{p_{q,x}^{(r)}}-p_{x,min}}{{d_{x}}}\right)}}{\eta' \left({{{{\bf{\bar p}}}^{(r)}}} \right)}+\frac{1}{2}\right]+p_{x,min},\\
		\label{P-q_y_est_final_fin}
		{{\hat p}_{q,y}}& = {d_{y}}\left[\frac{1}{2\pi}\angle \;\frac{1}{R}\sum\limits_{r = 1}^R {e^{j2\pi\left(-\frac{1}{2}+\frac{{p_{q,y}^{(r)}}-p_{y,min}}{{d_{y}}}\right)}}{\eta' \left({{{{\bf{\bar p}}}^{(r)}}} \right)}+\frac{1}{2}\right]+p_{y,min}.
	\end{align}
\end{subequations}

Note that with linear interpolation technique, the position realizations are not confined to be on grid points. Hence, the proposed IS-DPD does not suffer from
the serious off-grid problem.
\subsection{Complexity analysis}
This part evaluates the complexity of the proposed IS-based estimator. The complexity is counted through the order of the number of complex-valued multiply operations. Most of the computational resource is consumed by obtaining the
discretized importance function, generating realizations and evaluating the weighting coefficient. According to the definition of ${\bar {\cal L} _q}\left( {{p_{{n_x}}},{p_{{n_y}}}} \right) $, for one emitter the complexity for obtaining the	discretized importance function is $O\left( {N_x}{N_y}KLMN^2   \right)$. As there are Q emitters, each emitter have $R$ realization pairs, recalling STEP \ref{step_mid}-STEP \ref{step_ed}, the complexity of realization generation is $O\left(Q{N_x}+ QR{N_y}\right)$. \eqref{E_eta_another_explict} involves the matrix inverse, whose computational complexity
is $O\left(QRKL\left(Q^2+MN^2\right)   \right)$. So the total computational complexity of the IS-DPD is $O(Q{N_x}{N_y}KLMN^2+QR{N_y}+QRKL(Q^2+MN^2))$. On the other hand, the computational complexity of the exhaustive ML grid search algorithm is $O((N_xN_y)^QKL(Q^3+QMN^2))$, it can be seen that when $(N_xN_y)^Q\gg R$, which is commmon in the
moving receiver scenario, the new IS-based ML DPD estimator
exhibits remarkable computational savings.  These observation are summarized in Table \ref{table_example}.  The relative processing time (Rel. Proc. Time) with respect to the proposed method is obtained in a scenario with moving receivers as Fig. \ref{scenario}, where $R=1000,Q=2,N_x=100$ and $N_y=100$.
\begin{table}[!t]
	\renewcommand{\arraystretch}{1.3}
	\caption{Complexity Assessment of the Considered DPD Algorithms}
	\label{table_example}
	\centering
	\begin{tabular}{||c||c|c||}
		\hline\hline
		Algorithm & Complexity & Rel. Proc. Time \\
		\hline
		IS-DPD& $O(Q{N_x}{N_y}KLMN^2+QR{N_y}+QRKL(Q^2+MN^2))$ & 1\\
		\hline
		Exhaustive ML search & $O((N_xN_y)^QKL(Q^3+QMN^2))$ & 9091\\
		\hline
	\end{tabular}
\end{table}
\section{Simulation results}
\label{sec:Simulation}
To evaluate the performance of the proposed IS-based DPD approach,  we compare it with the AP-DPD algorithm \cite{oispuu2010direct} that is one of the iterative implementations of ML-type methods, the SML-DPD and MVDR-DPD proposed in \cite{tirer2017high}, which are beamforming-based methods. The position estimation performance is evaluated in terms of the root mean square error (RMSE)
\begin{equation}
	\label{RMSE}
	{\rm{RMSE}} = \sqrt {\frac{1}{{{N_{Mc}}}}\sum\nolimits_{{n_{Mc}} = 1}^{{N_{Mc}}} {{{\left\| {{\bf{\hat p}}_q^{\left[ {{n_{Mc}}} \right]} - {{\bf{p}}_q}} \right\|}^2}} },
\end{equation}
where $N_{Mc}$ is the number of ensemble runs for each test point. ${\bf{\hat p}}_q^{\left[ {{n_{Mc}}} \right]}$ is the estimated $q$-th emitter position at the $n_{Mc}$-th trial. The proposed method and aforementioned algorithms are compared with the Cam\'er-Rao lower bound (CRLB) \cite{qin2018decoupled}, which reflects the theoretical achievable performance taken as a benchmark for all  considered algorithms.
\begin{figure}[!t]
	\centering
	\includegraphics[width=1\linewidth]{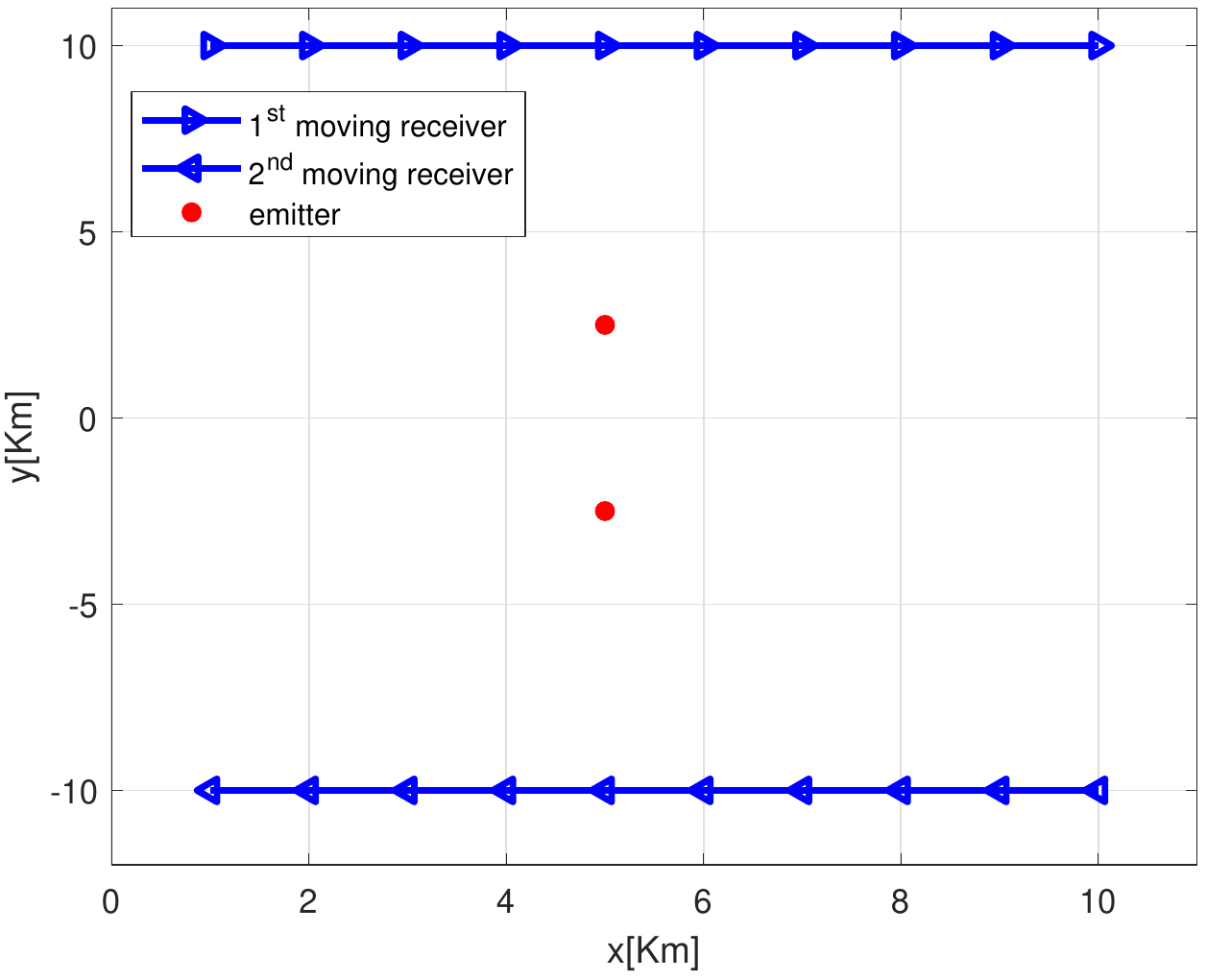}
	\caption{The simulation scenario.}
	\label{scenario}
\end{figure}

Consider two moving receivers equipped with ULA consisting of $M=3$ half-wavelength spaced elements. They move from $\left[ {1,10} \right]$ Km to $\left[ {10,10} \right]$ Km and $\left[ {10,-10} \right]$ Km to $\left[ {1,-10} \right]$ Km, respectively. The receivers intercept emitted signals from interested emitters once moving 1 Km each, thus $K=10$ in this case. The scenario is depicted as Fig. \ref{scenario}. The length of the observation time interval is $T=12.8$ ms and the receivers' speed is 300 m/s. The orientation of the array is the same as the moving direction of the corresponding receiver. Emitters are located in a square area of $10 \times 20$ Km $\times$ Km, transmitting flat narrowband Gaussian signals with equal power and bandwidth $B=10$ KHz, the number of emitters is a priori knowledge. The signal carrier frequency is ${f_c}=0.1$ GHz, and the signal propagation speed is set as $c=3\cdot10^8$ m/s. The waveform of transmitted signals is known. The down-converted signal is sampled at 10 KHz (by complex sampling) in each interception interval, i.e., each test uses $N=128$ samples. For each emitter, the channel attenuation is randomly generated following the normal distribution with mean one and standard deviation 0.1, and the channel phase is selected from $\left[ 0,2\pi \right]$ uniformly.  The ensemble runs is $N_{Mc}=5000$.

\begin{figure}[!t]
	\centering
	\includegraphics[width=0.6\linewidth]{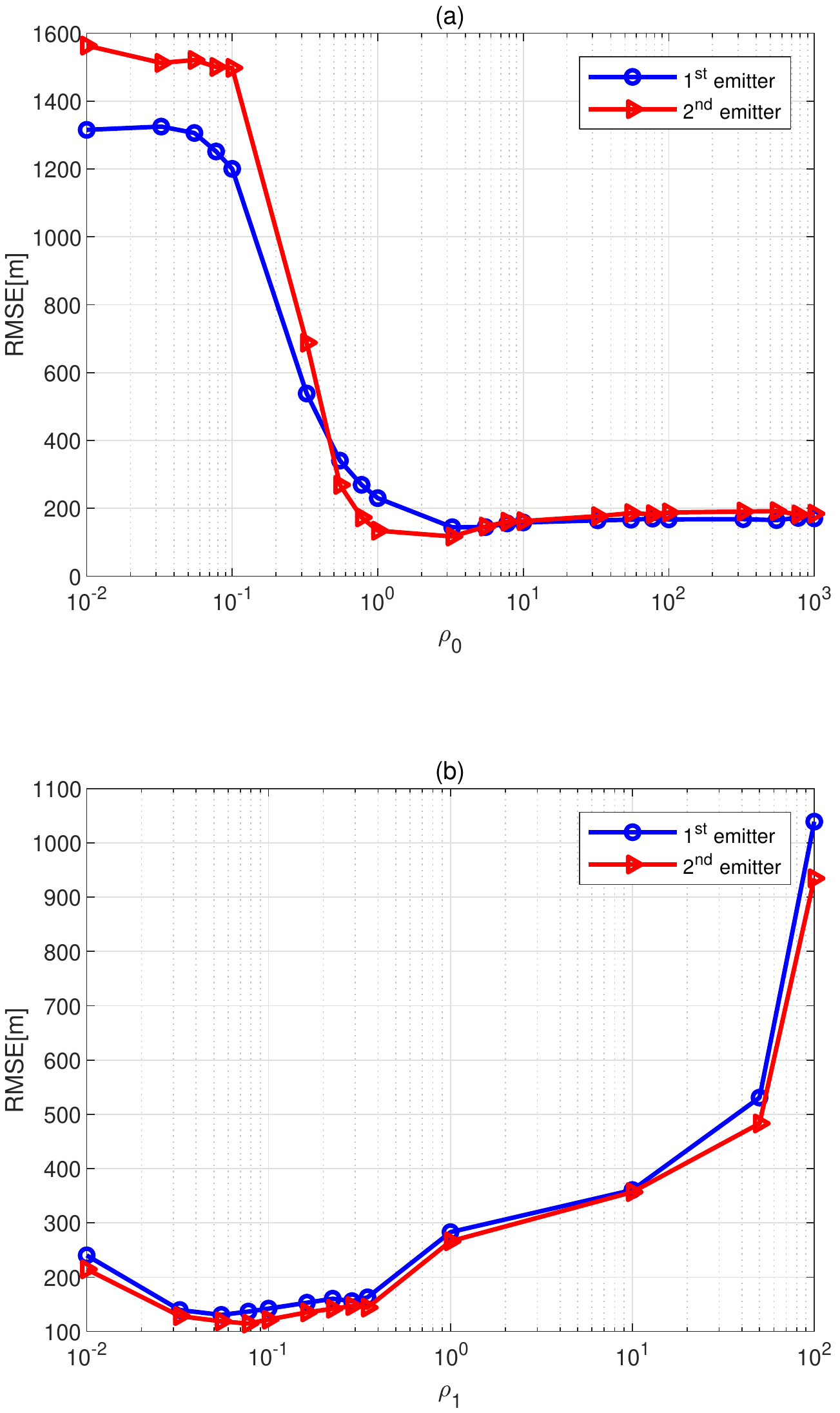}
	\caption{Estimation performance vs. (a) ${\rho _0}$ and (b) ${\rho _1}$.}
	\label{fig_ro0_ro1}
\end{figure}

As mentioned in Section \ref{subsec:On the use of importance sampling function},  ${\rho _0}$ and ${\rho _1}$ are design parameters that should be
carefully chosen. To investigate the effect of each parameter on
the estimation performance, we vary one of them and fix another one. The scenario contains two emitters located at $\left[5,2.5 \right]$ Km and $\left[5,-2.5 \right]$ Km . The rest parameters are fixed at $R=1000$, $\Delta _{p_x}=1$ Km and $\Delta _{p_y}=1$ Km. Fig. \ref{fig_ro0_ro1} shows the performance of the IS-based DPD estimator at SNR=25 dB versus (vs.)  ${\rho _0}$ and  ${\rho _1}$. When the value of ${\rho _0}$ is small, the estimator exhibits very poor estimation performance as seen in Fig. \ref{fig_ro0_ro1}a. Increasing ${\rho _0}$ can remarkably improve the estimation
accuracy. This is reasonable because it can be easily
inferred from the infinite limit involved in Pincus’ theorem. On the other hand, as illustrated in Fig. \ref{fig_ro0_ro1}b, the optimal value of ${\rho _1}$ is between 0.005 and 0.1, since this design parameter is used to control the spans of the main lobes of ${\bar {\cal L} _{p_{q,x}}}\left( {p_x} \right)$ and ${\bar {\cal L} _{{p_{q,y}}\left| {{p_{q,x}}} \right.}}\left( {{p_{{y}}}\left| p_x\right.} \right)$. Small ${\rho _1}$ may not neutralize the effect of the additive noise, while a too large value renders the main lobes be extremely narrow, leading to the true position always lies outside. Large estimation bias is inevitable in both aforementioned conditions. In the following simulations, ${\rho _0}$ and ${\rho _1}$ are set as 100 and
0.035, respectively.

\begin{figure}[!t]
	\centering
	\includegraphics[width=0.6\linewidth]{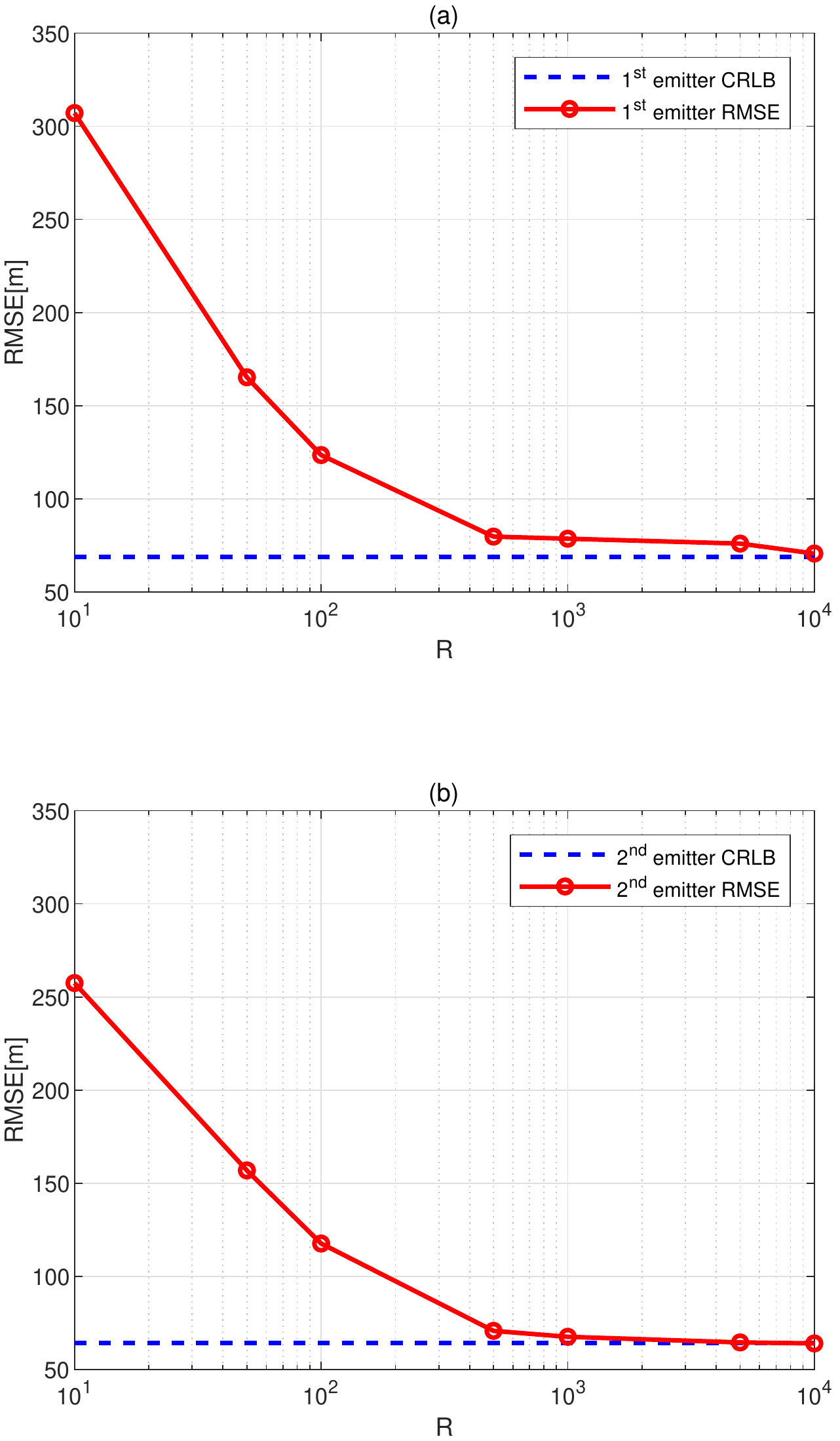}
	\caption{Estimation performance vs. ${R}$ for (a) the 1st emitter and (b) the 2nd emitter.}
	\label{fig_R}
\end{figure}

The next experiment evaluates the impact of the parameter $R$ on the performance of the proposed DPD estimator at SNR=5 dB. Other parameters remain the same as above. The results in Fig. \ref{fig_R} match well with the well-known estimation theory that a large enough value of $R$ contributes to the consistent mean estimation for \eqref{P-q_x_est_final_fin} and \eqref{P-q_y_est_final_fin}. However, the computational load inevitably increases as $R$ becomes larger. Using a relatively small value such as $R=1000$ yields a satisfactory trade-off between performance and complexity, since the accuracy increases insignificantly when $R$ is larger than 1000. In the following simulations, ${R}$ is fixed at 1000.

\begin{figure}[!t]
	\centering
	\includegraphics[width=0.6\linewidth]{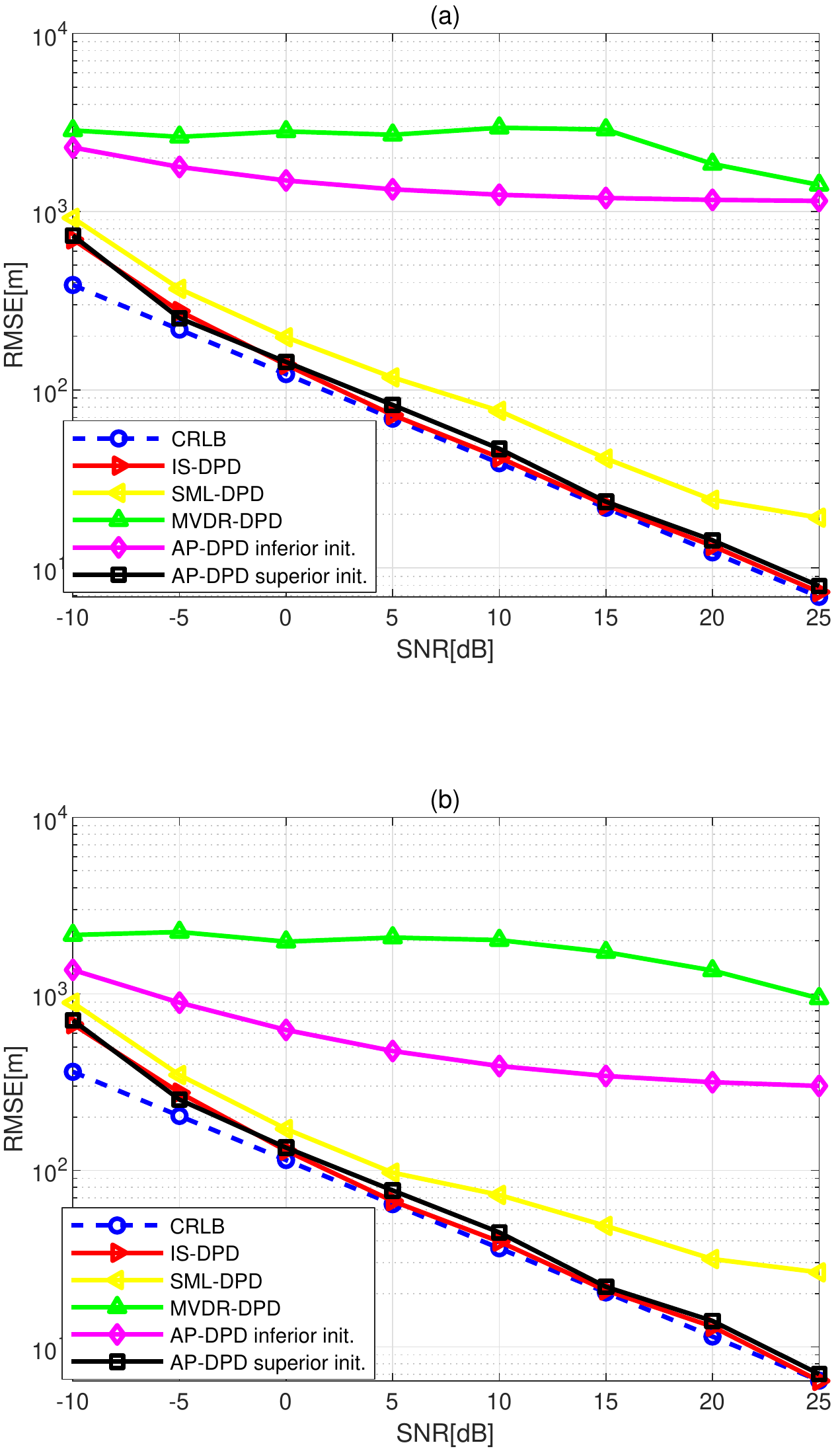}
	\caption{Estimation performance of different algorithms vs. SNR for (a) the 1st emitter and (b) the 2nd emitter.}
	\label{fig_SNR}
\end{figure}

Fig. \ref{fig_SNR} presents the RMSE of position estimation for varying SNR. The MVDR-DPD estimator passes the received signals through $LM$ digital Chebyshev type I filters and uses the resulted outputs as snapshots. The peak-to-peak ripple of the set of filters is 0.5 dB. The order of the first and the last filters is 3, while the rest are of order 6. The passband of the $n_f$-th filter (normalized by bandwidth $B$) is given by 
\begin{equation}
	\label{filter}
	\left[0.3\frac{n_f-1}{LM-1},0.7+0.3\frac{n_f-1}{LM-1}\right],n_f=1,2,\ldots,LM. 
\end{equation}
And for the AP-DPD algorithm, we consider two cases that the initial position of the iterative process is far from and close to the true emitter position. Fig. \ref{fig_SNR} shows the estimation accuracy of the proposed IS-based DPD estimator is very close to the CRLB. The AP-DPD with a good initialization can achieve similar performance as IS-DPD. However, when the initial guess is far away from the true value,  the performance of AP-DPD deteriorates significantly. The MVDR-DPD fails in resolving emitters thus it performs poorly. SML-DPD is inferior to IS-DPD because the correlation among emitters is neglected. The simulation results validate that the IS-DPD is a  more robust and accurate
implementation of the ML estimator.

\begin{figure}[!t]
	\centering
	\includegraphics[width=0.6\linewidth]{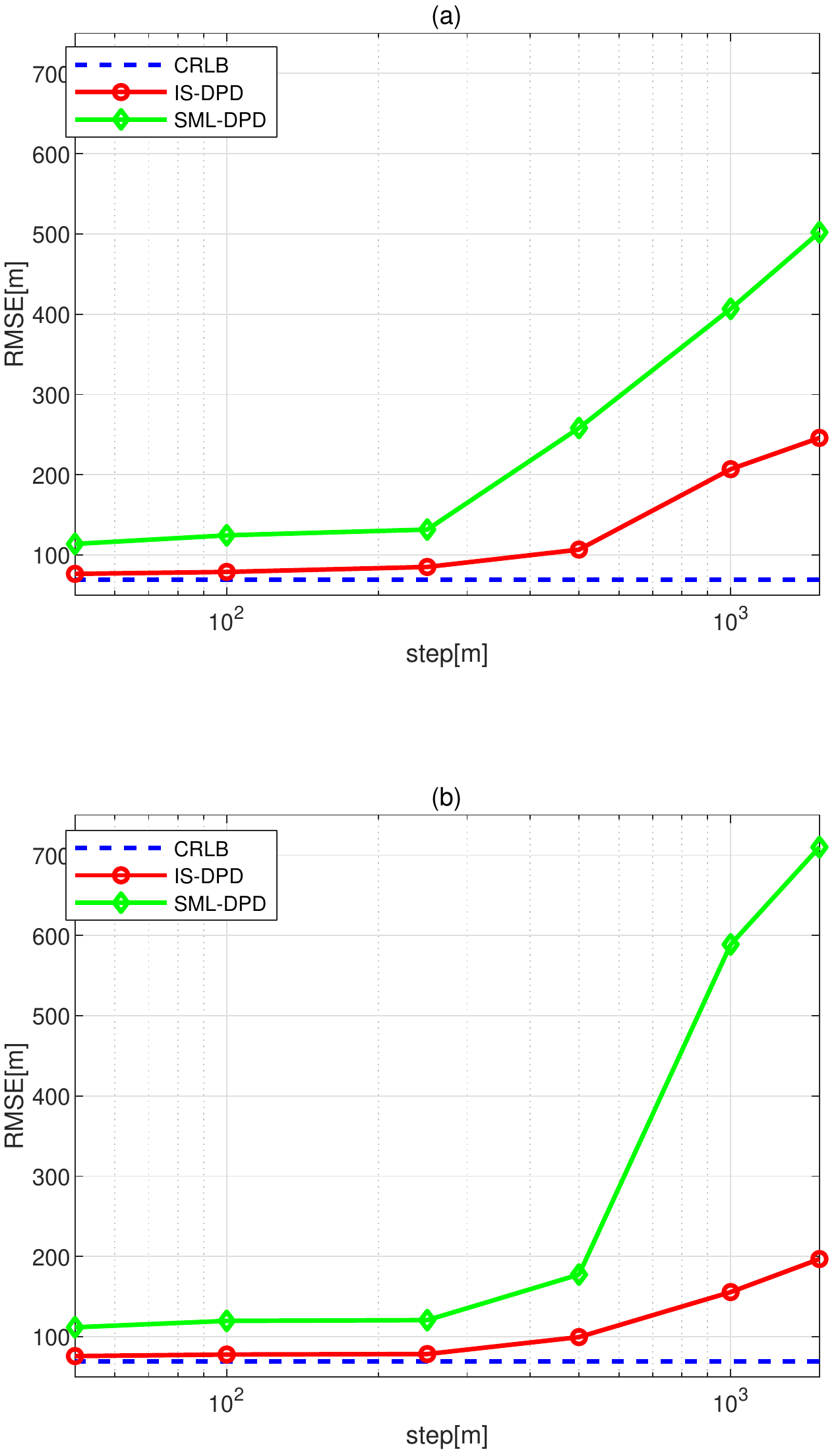}
	\caption{The RMSE of each estimator vs. search grid step for (a) the 1st emitter and (b) the 2nd emitter.}
	\label{fig_step}
\end{figure}

So far, comparisons have been performed versus SNR. To study the influence of the grid density on the performance of different estimators, we vary the grid step with SNR fixed at 5 dB, the true locations of two emitters are unchanged. The SML-DPD is compared as one representative example of
the grid-based algorithm. It can be seen
from Fig. \ref{fig_step} that, as the grid
step gets larger, the estimate becomes less
accurate for all methods. Nevertheless, the estimation performance of IS-DPD is still superior to that of SML-DPD. Apparently different from the proposed IS-based estimator, SML-DPD is restricted by the grid step. The results in Fig. \ref{fig_step} match well with the theoretical results in
Section \ref{sec:Introduction} and Section \ref{sec:Implementation details}, which shows the benefit due to the utilization of linear interpolation in proposed DPD estimator. Advancing the other DPD methods, the off-grid problem in IS-DPD can be considerably mitigated.

\begin{figure}[!t]
	\centering
	\includegraphics[width=0.6\linewidth]{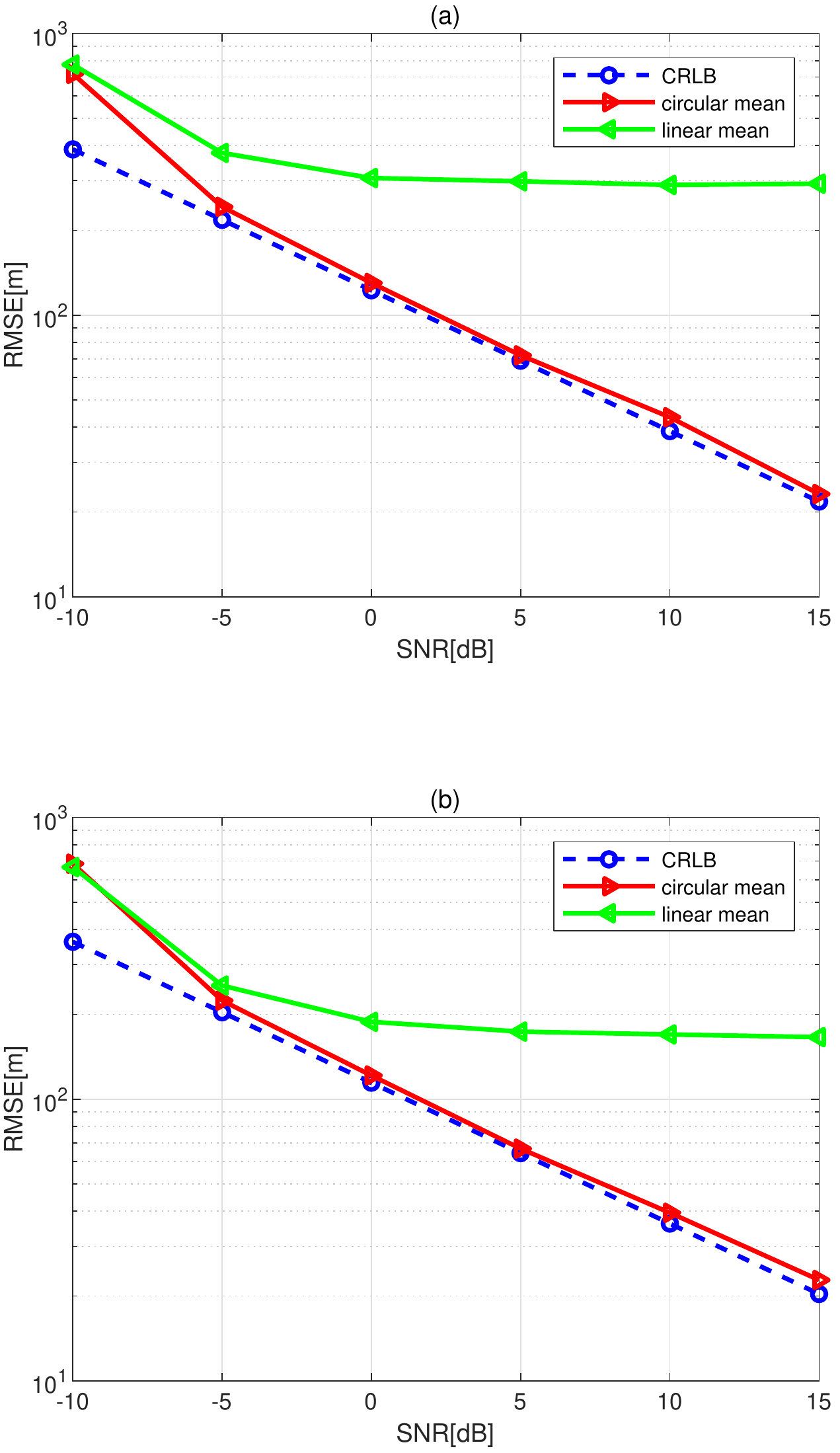}
	\caption{Estimation performance vs. the way for obtaining sample mean in proposed IS-DPD algorithm for (a) the 1st emitter and (b) the 2nd emitter.}
	\label{fig_mean}
\end{figure}

In the last experiment, we focus on the impact of the way for estimating sample mean in the proposed IS-DPD algorithm. Fig.  \ref{fig_mean} shows the RMSE of position estimation with linear mean and circular mean vs. SNR. Applying the linear mean of samples in the IS-DPD results in a large estimation bias, thus it can be observed that the performance deviates from the CRLB, and the gap between the RMSE and CRLB becomes larger as the SNR increases. The substitution of the linear mean by circular mean significantly reduces the estimation bias and contributes to RMSE achieving the CRLB level. The simulation results validate the advantage of the proposed circular mean based IS-DPD algorithm.

\section{Conclusion}
\label{sec:Conclusion}
This paper discusses location estimation of multiple stationary narrowband radio-frequency emitters based on 
	angle and Doppler shift measurements, where the transmitted signals are observed by multiple moving array receivers.
We developed a new implementation of DPD ML estimation, as the number of emitters and the transmitted signal waveform are known a priori. Based on the importance sampling concept, the computational complexity of the new ML DPD algorithm is significantly lower than the traditional multi-dimensional grid search methods, and the off-grid problem is alleviated. Moreover, the proposed method can be implemented separately and run in parallel. In addition, it does
not require initial parameter estimates but still
guarantees global optimality of the likelihood function. Simulation
results show the superiority of the proposed IS-DPD over the state-of-the-art DPD methods in both accuracy and robustness.
\section*{Acknowledgments}
This work was supported in part by the National Natural Science Foundation of China (NSFC) under Grant 61771108.
	\bibliography{IEEEabrv,mybibfile}
	
\end{document}